# Phase-resolved Detection of Ultrabroadband THz Pulses inside a Scanning Tunneling Microscope Junction

Melanie Müller[1*], Natalia Martín-Sabanés[1,2], Tobias Kampfrath[1,2], and Martin Wolf[1]

[1]*Department of Physical Chemistry, Fritz-Haber Institute of the Max-Planck Society, Faradayweg 4-6, 14195 Berlin, Germany.*

[2]*Department of Physics, Freie Universität Berlin, Arnimallee 14, 14195 Berlin, Germany.*

*Corresponding author: m.mueller@fhi-berlin.mpg.de

Coupling phase-stable single-cycle terahertz (THz) pulses to scanning tunneling microscope (STM) junctions enables spatio-temporal imaging with femtosecond temporal and Ångstrom spatial resolution. The time resolution achieved in such THz-gated STM is ultimately limited by the sub-cycle temporal variation of the tip-enhanced THz field acting as an ultrafast voltage pulse, and hence by the ability to feed high-frequency, broadband THz pulses into the junction. Here, we report on the coupling of ultrabroadband (1-30 THz) single-cycle THz pulses from a spintronic THz emitter (STE) into a metallic STM junction. We demonstrate broadband phase-resolved detection of the THz voltage transient directly in the STM junction via THz-field-induced modulation of ultrafast photocurrents. Comparison to the unperturbed far-field THz waveform reveals the antenna response of the STM tip. Despite tip-induced low-pass filtering, frequencies up to 15 THz can be detected in the tip-enhanced near-field, resulting in THz transients with a half-cycle period of 115 fs. We further demonstrate simple polarity control of the THz bias via the STE magnetization, and show that up to 2 V THz bias at 1 MHz repetition rate can be achieved in the current setup. Finally, we find a nearly constant THz voltage and waveform over a wide range of tip-sample distances, which by comparison to numerical simulations confirms the quasi-static nature of the THz pulses. Our results demonstrate the suitability of spintronic THz emitters for ultrafast THz-STM with unprecedented bandwidth of the THz bias, and provide insight into the femtosecond response of defined nanoscale junctions.

**KEYWORS: scanning tunneling microscopy, THz voltage sampling, spintronic THz emitter, broadband THz pulses, tip antenna response, ultrafast photocurrents**

THz-gated scanning tunneling microscopy (THz-STM) combines Ångstrom spatial with femtosecond temporal resolution, which has been impressively demonstrated on single molecules[1] and semiconductor surfaces[2]. Following the original idea of junction-mixing STM[3,4], the concept of THz-STM is based on the rectifying nature of an STM junction exhibiting nonlinear I-V characteristics, leading to a net DC current upon modulation of the junction bias with ultrafast voltage pulses. Adapting this concept, THz-STM utilizes 'wireless' free-space coupling of ultrafast voltage pulses to the STM by illumination with coherent broadband THz radiation[1,2,5–8]. The STM tip hereby acts as a broadband antenna strongly enhancing the quasi-static THz electric field, allowing for the application of large sub-picosecond bias pulses at moderate incident THz field strength. In contrast to ultrafast THz scanning near-field optical microscopy (THz-SNOM), where the scattered THz near-field yields information about the local transient dielectric response of the sample[9], THz-STM demands that the antenna-enhanced THz field exhibits single-cycle character in order to ensure unambiguous control of the tunneling current within less than one optical cycle of the THz bias. Therefore, optimizing THz-STM operation requires precise knowledge of the THz voltage transient across the junction and hence broadband characterization of the tip antenna response directly in the STM environment.

The coupling of THz radiation to scanning probe tips has been widely studied in the context of apertureless THz-SNOM[10–17]. It is known that metallic tips act as long wire antennas,[13,14] which low-pass filter the broadband incident THz radiation[11] and exhibit highly directionally emitting and receiving properties[14]. The STM tip-enhanced THz waveform will, thus, differ considerably from the incident THz waveform, depending on the specific STM geometry, the mesoscopic shape of the tip wire, as well as the incident THz spectrum. Yet, detailed experimental characterization of the STM tip antenna properties and its effect on the THz waveform inside the STM environment remain scarce[7,8]. In particular, understanding the antenna response of the STM tip over a broad frequency range exceeding 10 THz will be crucial to extend the concept of THz-STM towards higher frequencies up to the multi-THz[9], mid-infrared[18–20], and even optical regime[21], potentially increasing the time resolution achievable in light-wave driven STM. To reach this goal, high-frequency single-cycle voltage transients have to be coupled into the STM with sufficient amplitude at MHz repetition rates, and reliable detection methods applicable to common STM setups are required for their characterization.

Recently, ultrabroadband single-cycle THz pulses were successfully generated from a metallic spintronic THz emitter (STE) with spectra covering the frequency range up to 30 THz without gap[22,23]. The spectral bandwidth of the STE output is determined by the duration of the incident pump pulse and the resulting carrier dynamics in the STE[24]. Due to its extremely large bandwidth, its fast single-cycle transients, and due to convenient THz polarity switching and polarization control via the STE magnetization, the STE is very well suited for THz-STM operation at high THz frequencies. In addition, it exhibits several additional advantages such as flexibility regarding pump photon energy, pulse duration, and excitation geometry, its easy handling as well as low cost[25]. Although the conversion efficiency of the STE is considerably lower compared to standard THz sources such as $LiNbO_3$[26,27] or photoconductive antennas[28,29], its high beam quality in combination with high THz bandwidth allows for tight focusing, facilitating peak field strength of 300 kV/cm at few mJ pump pulse energies[23]. Hence, high electric field strengths are achieved at comparably low THz power, making the STE an attractive THz source in particular for field-driven applications requiring high repetition rates such as THz-STM.

Here, we report on the generation and local detection of ultrabroadband single-cycle THz voltage pulses inside a metallic STM junction. We demonstrate that THz voltage transients with frequency components up to 15 THz and amplitudes up to 2 V can be achieved at moderate pump pulse energies of a few microjoules at 1 MHz repetition rate. To characterize the bandwidth, phase and voltage amplitude of the tip-enhanced THz field we sample its waveform directly in the time domain by THz-induced modulation of ultrafast photocurrents excited with near-infrared (NIR) femtosecond laser pulses in the tip-sample junction[8,30], as sketched in **Figures 1a)** and **1b)**. By comparison to the free-space THz waveform measured via electro-optic sampling (EOS), we can experimentally determine the receiving antenna response of the STM tip. Special care has to be taken to obtain the unperturbed tip-enhanced THz waveform without distortion by photoelectron dynamics, which is particularly crucial at the broad bandwidth employed here. In contrast to common near-field detection techniques, this sampling method does not rely on scattering methods requiring demodulation techniques and a vibrating tip, which is not available in common STM setups. Also, it provides direct access to the THz voltage as the relevant quantity for STM, rather than the electric field. Finally, we analyze the distance dependence from the few nanometer to the μm range, revealing a nearly constant THz voltage in line with the quasi-static nature of electromagnetic radiation when applied to sub-wavelength dimensions. Our results demonstrate the suitability of spintronic THz emitters as broadband source for the application of ultrafast voltage transients in STM, and highlight the importance of the tip antenna response that exhibits significant low-pass filtering and reduction of the THz bandwidth available for THz-STM operation.

**Experimental details.** The STE is excited with broadband 800 nm NIR pulses of 8 fs duration at 1 MHz repetition rate and under normal incidence, leading to THz pulse emission collinear with the NIR laser beam. The STE magnetization is controlled by a permanent magnet. The STE position along the focused NIR pump beam can be set by a translation stage for variation of the NIR spot size and fluence incident on the STE. The THz pulses are focused onto the STM tip at 68° incident angle with respect to the tip axis via an off-axis parabolic mirror (35 mm focal length) integrated on the STM platform inside the ultrahigh vacuum (UHV) chamber. A second parabolic mirror is used to focus 8 fs NIR laser pulses with a pulse energy of $E_{\mathrm{p,STM}} = 0.6$ nJ into the STM (NIR spot size ~ 6 µm) for photoexcitation of the junction at a variable time delay $\Delta t$ compared to the arrival time of the THz pulses. Both THz and NIR pulses are polarized along the tip axis. All experiments are performed at room temperature and under UHV (pressure 1e-10 mbar) conditions.

**THz-induced photocurrent modulation. Figure 1c)** shows the photocurrent-voltage characteristics of the photoexcited junction at a tip-sample distance of $d = 1$ µm in the absence of the THz pulse. As the NIR focus is larger than the tip-sample distance, photoelectrons are excited both from tip and sample and the photocurrent $I_{\mathrm{ph}}$ reverses sign at small negative DC bias. At large positive (negative) DC bias, as sketched in **Figure 1b)**, the photocurrent is dominated by pure photoelectron emission from the tip (sample), whereas in the low bias regime, it is composed of both tip and sample photoelectrons depending on their respective energy distributions. At large distances, photoemission is expected to occur either in the multiphoton regime[31–33] or optical field-driven regime[34–36] depending on the applied NIR field strength and classified by the Keldysh parameter[37]. With our experimental parameters, we obtain a Keldysh parameter in the range of $\gamma \sim 10\text{-}20$. Hence, we clearly operate in the multiphoton photoemission regime, as confirmed by the scaling of the power dependence of the photocurrent shown in the inset in Figure 1c). We note that transient thermionic emission, i.e, the emission of thermalized hot electrons, can also contribute to the photocurrent, though we expect this to be insignificant at our conditions[32,38,39]. At high DC bias and, in particular, at short distances, photo-assisted tunneling through the narrowed barrier[40] may also contribute to or even dominate the photocurrent.

Applying a THz pulse to the STM junction in addition to the static DC bias leads to a change in the laser-induced photocurrent due to the THz field[8,30]. As the time scale of photoemission is given by the temporal envelope of the 8 fs NIR laser pulse and occurs thus nearly instantaneous on the time scale of the THz pulse, the photoemission process is sensitive only to the *instantaneous* THz field at a given NIR-to-THz pulse time delay $\Delta t$. One may, thus, make the quasi-static approximation which assumes that the THz-induced change in photocurrent will be determined by the local slope of the static $I_{ph}$-V characteristics, as sketched in **Figure 1c)**. As will be demonstrated below, our measurements corroborate this assumption provided that non-instantaneous effects are eliminated. In the case of a linear $I_{ph}$-V curve with local slope $m_{I-V}$, as obtained at high positive bias in **Figure 1c),** the instantaneous THz voltage $U_{\text{THz}}(\Delta t)$ can be obtained from the THz-induced change in photocurrent $\Delta I_{\text{THz}}$ simply by $U_{\text{THz}}(\Delta t) = \Delta I_{\text{THz}}(\Delta t)/m_{I-V}$. Varying the time delay $\Delta t$, thus, directly yields the calibrated THz voltage transient as plotted in **Figure 1d)**, measured at 8 V DC bias and 1 μm gap distance. Reversing the sign of the STE magnetization flips the THz polarity,[22] and we find that the tip-enhanced THz waveform also reverses sign while maintaining its overall shape. Besides demonstrating this strikingly simple THz bias control in the STM provided by the STE magnetization, this finding confirms the interpretation of the employed sampling technique.

At this point, it is important to emphasize that non-instantaneous effects such as photoelectron propagation in the junction and long lifetimes of hot electrons inside the photoexcited tip and sample surface could alter the measured waveform and, thus, have to be ruled out. The lifetime of hot electrons in metals[41] is in the range of few fs to several tens of fs depending on the sampled energy window and temperature. At large gap distances, and in the absence of strong transient thermionic emission, lifetime effects can be neglected because the photoelectrons are emitted predominantly above the barrier, and tunneling can occur only at the very barrier top, where extremely short lifetimes of few fs are expected. Furthermore, THz streaking of photoelectrons back into the tip/sample[30,42] could also result in a modified waveform, but can be precluded by operating at high DC bias and low THz field strength (smaller than the DC field). The latter is only strictly given in the single electron regime, where the electron trajectories are solely determined by the combined DC and THz field. At larger photocurrents with many hundreds to thousands of electrons per pulse, space charge dynamics has to be considered and can significantly alter the measured THz waveform by acting as an effective low-pass filter. Likewise, the excitation of long-lived hot electrons in the confined tip volume might lead to increased cooling times and thus to broadening of the measured THz waveform at high excitation densities. Hence, care has to be taken to operate in a low-excitation limit at small photocurrents. We find that peak currents of 10 to few 100 electrons per pulse are usually sufficiently low to neglect broadening effects (see Supporting Information for more detailed information). It should be noted, however, that the critical photocurrent depends on the sharpness of the tip and the net accelerating forces acting on the photoelectrons, i.e., on the DC field as well as on the space charge density in the junction, given by the effective photoemission area and the tip-sample distance. To ensure instantaneous THz near-field sampling within the quasi-static approximation, we ,thus, verify that the sampled THz waveform is insensitive to the THz polarity, incident THz field strength, DC bias and tip-sample distance (details are discussed in the Supporting Information and in Figure 5 below).

**Tip-antenna response function.** Having established a reliable method to sample the THz waveform inside the STM junction, we are now able to characterize the tip antenna response by comparison of the unperturbed incident THz field and the received THz voltage transient. **Figure 2a)** shows the THz electric field $E_{in}$ obtained by electro-optic sampling using a 300 µm thick ZnTe(110) detection crystal in an identical reference beam path and after deconvolution with the detector response[43]. Noise-like high-frequency features arise from the deconvolution process because the electro-optic sampling transfer function of our crystal has a relatively low amplitude at frequencies above 5 THz. Therefore, the electro-optic signal at these frequencies has lower signal-to-noise ratio. The corresponding THz voltage transient in the STM is shown in **Figure 2b)**. The low-pass filter effect of the tip antenna becomes immediately obvious in the time domain data, where a half-cycle peak-to-peak separation of 115 fs is attained in the tip-enhanced waveform, noting the maximum speed at which the THz bias can be reversed in the STM. In the time domain, the THz near-field in the STM junction $E_{STM}(t)$ is given by convolution of the incident THz field $E_{in}(t)$ with the tip impulse response. More convenient access to the tip antenna response is obtained by Fourier transformation of the time domain data into the frequency domain, where the tip-enhanced THz field is simply the product of the incident THz field and the complex tip transfer function $H(\omega)$. Equivalent to a receiving antenna's output voltage and following wideband antenna theory,[44] we can also use the experimentally accessible frequency-dependent THz voltage $U_{\text{THz}}(\omega)$ instead of the tip-enhanced THz field and relate it to the incident THz electric field $E_{\text{in}}(\omega)$ by

$$U_{\text{THz}}(\omega) = H(\omega)E_{\text{in}}(\omega). \tag{1}$$

Here, $H(\omega)$ is the receiving antenna transfer function whose dimension is meters. It depends on the angle $\theta$ of incidence of the THz beam[14], the active antenna dimensions given by the thickness and illuminated length of the STM wire[11], and geometric details of the tip[7]. The Fourier spectra of the unperturbed and tip-enhanced waveforms are plotted in **Figure 2c)**, revealing a pronounced bandwidth reduction and simultaneous shift of the tip-enhanced THz spectrum to lower frequencies. The tip transfer function $H(\omega)$ is then simply obtained by dividing the two complex Fourier spectra, and its amplitude and phase are plotted **Figure 2d)**. The amplitude $|H|$ reflects the coupling efficiency of the tip antenna at a specific THz frequency, and its phase describes pulse distortions induced by the tip.

We model the transfer function by a simple *RLC* circuit with resistance $R$, capacitance $C$ and inductance $L$ connected in series[11,30] (see Supporting Information for more details). The circuit parameters are adjusted to best reproduce the THz voltage transient from the incident waveform. We obtain reasonably good agreement for $R = 300\ \Omega$, $L = 0.32$ nH and $C = 35$ fF, yielding a resonance frequency of $f_0 = 1/\left(2\pi\sqrt{LC}\right) = 0.05$ THz. The *RLC*-filtered waveform and the *RLC* transfer function are also plotted in **Figures 2b)** and **2d)**. We note that the resonance is not visible here as it lies at very low frequencies outside our THz spectrum. From the *RLC* model it becomes clear that the STM tip exhibits mostly inductive behavior in the frequency range of the STE, causing the observed low-pass filter effect. The flat spectral phase implies that no significant distortion (chirp) of the THz pulse is introduced by the tip due to the absence of group delay dispersion in the frequency range of the STE. Deviations of the measured and *RLC*-filtered waveforms in **Figure 2b)** are also clearly visible, indicating the limitations of the *RLC* model to describe our results. In fact, the active antenna length, i.e., the inductance and capacitance "seen" by the THz field, will depend on the THz spot size which varies with frequency, which is not accounted for in the model. Moreover, scattering and retardation, as introduced by geometrical details such as the tip shaft, cannot be predicted by the *RLC* model. Strategies to enhance the high THz frequencies in the STM junction, thus, should include decreasing the inductance of the STM wire and modification of the tip transfer function via resonant tip shaping[7]. We emphasize that due to the steep increase of the THz field enhancement at low frequencies, efficient reduction of the THz near-field duration requires active suppression of the low THz frequencies, which would otherwise dominate the response. This could be achieved by spectral matching of the incident THz spectrum reciprocal to the STM tip response. Finally, we note that additional low-pass filtering is caused by the frequency-dependent orientation of the antenna lobes[14], leading to less efficient coupling of the high THz frequencies at large incident angles with respect to the tip axis.

To demonstrate the usefulness of the transfer function, we apply a phase shift to the incident THz field by moving the STE inside the convergent beam of the NIR pump laser, as sketched in **Figure 3a)**. In such a curved-wavefront excitation scheme the local radius of curvature of the NIR beam is imprinted on the generated THz field.[28] Upon propagation to the far-field, the emitted THz pulse, thus, acquires a frequency-dependent phase shift that depends on the STE position. In particular, an intermediate THz focus is expected when placing the STE in the convergent part of the NIR beam.[28] As seen in the top panel in **Figure 3b)**, we observe a transformation from a rather symmetric to a more asymmetric pulse shape of the tip-enhanced THz waveform when moving the STE further away from the NIR focus. We then apply the measured tip transfer function to the phase-shifted incident THz fields and compare the reconstructed antenna-enhanced waveforms to the corresponding THz waveforms measured in the STM. Specifically, we multiply the complex Fourier spectra $E_{\text{in}}(\omega)$ of the deconvoluted THz waveforms obtained from EOS (see Figure S3 in the Supporting Information) with $H(\omega)$ to obtain the calculated Fourier spectra $U_{\text{rec}}(\omega)$ of the THz voltage received by the tip antenna. Inverse Fourier transformation of $U_{\text{rec}}(\omega)$ then yields the antenna-enhanced THz waveforms plotted in the bottom panel in **Figure 3b)**. Comparison of the waveforms for three different STE positions reveals that the small phase shifts apparent in the measured THz near-field waveforms are clearly reproduced by the respective calculated waveforms, demonstrating the validity of the experimentally obtained tip transfer function. This is advantageous for routine THz-STM operation using different THz waveforms, since once $H$ is known it allows to predict THz voltage transients in the STM from EOS reference data, given that the antenna geometry and effective length remains constant.

**THz amplitude scaling.** An important measure for the applicability of the STE for THz-STM is the achievable THz voltage in the STM junction. **Figure 4a)** shows the peak THz voltage versus NIR pump pulse energy for different STE positions at 1 MHz repetition rate. Corresponding THz waveforms are shown in **Figure 3b)** and **Figure 4b)**. We find that the THz voltage increases monotonically with the pump pulse energy. Peak THz voltages of up to 2 V can be achieved when placing the STE at $\Delta z_{STE} = $ 4-5 mm in front of the NIR focus ($\Delta z_{STE} = 0$). At distances closer to the NIR focus and, thus, smaller pump spot size, the achievable THz field strength is reduced. We assign this behavior to thermal heating of the STE resulting in a reduced STE magnetization and hence efficiency, as well as to the decreased THz emission efficiency of sub-wavelength volumes[45,46]. In addition, ablation and white light generation in the sapphire substrate prohibit the use of higher pump pulse energies at distances very close ($|\Delta z_{STE}| < 1$ mm) to the NIR focus in the current setup. At distances far away from the NIR focus, the THz voltage in turn decreases due to the increasing NIR pump spot size as expected from the linear fluence dependence[22] (see Supporting Information). At our laser parameters, using few-femtosecond laser pulses with microjoule energy at intermediate repetition rates of 1 MHz, the optimum excitation conditions are not yet fully explored and are subject to future work. We expect that higher THz voltages can be achieved by optimization of the STE excitation geometry, i.e., using a weakly focused or collimated NIR pump beam.[23]

As seen in **Figure 4b),** the polarity of the THz waveform reverses sign by moving the STE from the (i) far distant convergent to the (ii) far distant divergent side of the NIR pump beam. This behavior is not unexpected because in case (i), the THz beam behind the STE exhibits an intermediate focus in which the THz wave undergoes a Gouy-type phase shift of ~180°. In the region close to the NIR focus ($\Delta z_{STE} = 0$) the behavior of the amplitude and shape of the emitted waveforms becomes more complex. We observe that the THz spectrum becomes slightly wider and the polarity of the THz waveform changes. The slightly enhanced detected bandwidth probably arises from the fact that the NIR focal planes at $\Delta z_{STE} = 0$ and at the tip position are conjugated planes that are imaged onto each other. Detailed analysis of THz propagation under such conditions is, however, beyond the scope of this work, and further investigations are currently in progress. Our results, thus, demonstrate that high THz voltages can be achieved in the STM using the STE as THz source, but point out the importance of an optimized STE excitation geometry at microjoule-level pulse energies and intermediate repetition rates as required for THz-STM.

**Tip-sample distance dependence.** We finally analyze the dependence of the tip-enhanced THz waveform on the tip-sample distance. **Figure 5a)** shows the dependence of the current through the STM junction versus tip-sample distance $d_{rel}$ relative to the position of a setpoint of the NIR-illuminated junction of 200 pA current and 10 V DC bias. We estimate the actual gap distance at the setpoint to be 4.6 nm (see Supporting Information for more details). Upon retraction, the current drops rapidly within the first 1 nm. With increasing distance it stays nearly constant until at $d_{rel} \approx 600$ nm a pronounced photocurrent peak is observed presumably due to interference effects[47] of the exciting NIR laser pulse in the junction. Whereas at far gap distances multiphoton photoemission above the broad barrier dominates, see Figure 1b), tunneling of photoexcited electrons through the narrowed barrier is expected to contribute significantly at close gap distances[48]. At the setpoint distance, we most likely operate in a mixed regime of photoemission, photo-assisted tunneling and DC field emission. Disentangling the different contributions requires a detailed analysis of the photocurrent nonlinearities and the potential barrier in the STM junction[48,49]. THz near-field waveforms are recorded at different distances by retracting the sample by a defined step, and they are plotted in **Figure 5b).** The corresponding I-V curves for calibration of the THz voltage are plotted in **Figure 5e)**. Here, we note that the negative current at small positive biases arises from photoelectrons emitted with sufficient excess energy from the sample, whose contribution is enhanced at close distances as the sample moves into the center of the NIR focus. To avoid influence from drift especially at the setpoint distance, the feedback is temporarily switched on again between each THz-NIR time delay to reference the tip position. We find that the THz waveform does not change considerably over a wide range of tip-sample distances. We further observe a nearly constant THz voltage applied to the STM junction as plotted in **Figure 5c)**.

To better understand the scaling with tip-sample distance, we perform frequency-domain simulations of the tip-enhanced THz field in the junction (using the RF-module of COMSOL Multiphysics, details are described in the Supporting Information). The THz-induced potential difference $U_{\text{THz}}$ applied between tip and sample can then be found by line integration of the tip-enhanced THz electric field $\boldsymbol{E}_{\text{THz}}$ across the junction,

$$U_{\text{THz}}(\omega, d) = \int_{\text{tip}}^{\text{sample}} \boldsymbol{E}_{\text{THz}}(\omega, d) \cdot \mathrm{d}\boldsymbol{r} \ . \tag{2}$$

**Figure 5d)** shows the distance scaling of the normalized THz-induced potential difference between tip and sample (solid blue curves, left ordinate) and the peak THz electric field (dashed blue lines, right ordinate) as obtained from simulations for three THz frequencies.

As expected, the THz electric field at the tip strongly increases at shorter distances. At the same time, we find that the THz voltage stays approximately constant over the investigated distance range, confirming our experimental observation and being consistent with time-domain simulations reported in previous work[2].

These results imply that the THz electric field in the junction is quasi-static, i.e., the metal surfaces of the tip and sample are equipotential surfaces for the THz field. Consequently, the THz field exhibits the same spatial distribution and distance scaling as a static field generated by a constant electric potential difference applied between tip and sample. We confirm this notion by simulation of the DC electric field, which is also plotted (light-red solid curve) in **Figure 5d)**. We, thus, conclude that 1) retardation effects are negligible over the investigated length scale $L$, which holds if $L \ll \lambda$,[50] and that 2) tip and sample behave as perfect electric conductors. Conversely, the presence of retardation effects and/or significant field penetration into the metal due to the frequency-dependent dielectric response would lead to a distance-scaling of the THz field that deviates from the static case, and thus a distance-dependent THz voltage. In fact, we find that at 1 THz, the calculated potential difference changes by less than 1% over the entire range, whereas at 10 THz, the THz voltage reduces by about 8% from 1000 nm to 1 nm. We assign these deviations at higher THz frequencies and large gap distances to retardation, indicating the limitations of the quasi-static approach. Within our experimental error, our results, thus, support the view of the quasi-static nature of THz voltage pulses applied to nanoscale junctions, although limitations at frequencies above 10 THz need to be further investigated.

Our observation that the THz waveform does not change with tip-sample distance further implies that the transfer function $H$ has a rather insignificant dependence on the gap distance, in agreement with THz near-field simulations reported in previous work[7]. As the capacitance increases at short distances[15,51], $H$ should in general depend on the gap distance, and the antenna resonance should exhibit a red-shift at closer distances. Considering, however, that the relative increase of the capacitance is small and that the tip response is dominated by its large inductance, only marginal red-shifts of the resonance frequency are expected at the low frequency end of our THz spectrum, which will not have drastic effects on the THz near-field. The THz frequency response of a metallic STM junction is, thus, predominantly governed by the macroscopic wire dimensions and the THz focus size, defining the active antenna length.

Finally, we note that at very close gap distances the THz waveform might undergo additional low-pass filtering due to the increased lifetime of photoexcited electrons tunneling at lower energies through the junction. This effect does not seem to be significant at the conditions used here. Given that the undistorted waveform is precisely known, such carrier-induced waveform distortions will, in turn, allow one to study few-femtosecond photocarrier dynamics with far sub-cycle temporal resolution and with a spatial resolution given by the localization of the photoexcited current.

In summary, we demonstrated efficient coupling of ultrabroadband single-cycle THz pulses from a spintronic emitter (1-30 THz) to the junction of a scanning tunneling microscope. The STE is an attractive source for THz-STM operation, not only due its high bandwidth and fast field transients, but also due to its simple and potentially fast (tens of kHz) control of the THz polarity and polarization via the STE magnetization. We demonstrated that THz voltage transients with frequencies up to 15 THz reaching peak voltages up to 2 V can be coupled to the STM junction. Phase-resolved detection of the THz voltage transient over a wide frequency range is achieved by the THz-induced modulation of laser-excited photocurrents through the STM junction. This allows for broadband characterization of the tip transfer function directly in the STM environment. Our results show that the low-pass filtering characteristics of the STM tip antenna is a crucial parameter in the design of ultrabroadband THz-STM when aiming at increased time resolution by applying even faster THz voltage transients. Moreover, our experimental finding that the THz voltage remains constant over a wide range of tip-sample distances verifies the quasi-static nature of THz pulses coupled to the STM. We believe that faster voltage transients can be achieved in the STM by resonant enhancement of high THz frequencies via tip antenna design[17] or by blue-shifting the incident THz spectrum, e.g. via control of the frequency-dependent propagation of the THz beam emitted from the STE. Modified STE excitation conditions with a collimated pump beam of suitable diameter and a subsequent THz telescope are expected to further increase the THz field amplitude incident on the tip. Easier broadband characterization of the incident THz electric field will be employed by using thinner electro-optic detection crystals[52]. At higher (tens of THz) frequencies, we expect deviations from the quasi-static behavior due to retardation and the frequency-dependent material response. In this regard, our work provides a direct route towards the experimental characterization of the phase and amplitude of multi-THz voltage transients applied to an STM junction with few-femtosecond resolution.

**Methods**

Experiments are performed in an ultrahigh vacuum (UHV) system (base pressure of $< 5$ x $10^{-10}$ mbar) at room temperature. The STM (customized Unisoku USM-1400 with Nanonis SPM controller) is equipped with two off-axis parabolic mirrors (PM) integrated on the spring-loaded STM platform (1× bare Au and 1× protected Ag, 1" diameter, 35 mm focal length). The beams incident angles are 68° with respect to the tip axis. The THz beam enters the UHV chamber via a 500 µm thick diamond window and is focused by the Au mirror. NIR pulses are focused via the Ag PM for photoexcitation of the STM junction. The tip position is fixed and the sample is moved for coarse motion and scanning. The two PMs are motorized and can be moved in xyz-direction (Attocube GmbH) for precise focus adjustment on the tip apex. The DC bias is applied to the sample and the current is collected from the grounded tip. The current amplifier (Femto DLPCA) is operated at a gain of $10^9$ V/A at 1 kHz bandwidth. The THz-induced current is detected by chopping the THz excitation beam incident on the STE at 607 Hz and lock-in detection. The Ag(111) sample was cleaned by repeated cycles of Ar+ sputtering and annealing up to 670 K. STM tips are electrochemically etched from 300 µm thick polycrystalline tungsten wire and transferred to UHV immediately after etching.

A broadband OPCPA laser system (Venteon OPCPA, Laser Quantum) delivering 8 fs VIS-NIR laser pulses (800 nm center wavelength) with 3 µJ energy at 1 MHz repetition rate is used for THz generation and photoexcitation of the STM (2 µJ are available for the THz-STM setup). Part of the laser output is focused by a PM (focal length = 50 mm) for THz generation from the STE (5.8 nm thick W/CoFeB/Pt trilayer on 500 µm sapphire substrate[22]) at normal incidence. The emitted THz radiation is collected by a second PM (focal length = 50 mm) and a 500 µm thick silicon wafer is used to block the collinear NIR pump beam. The THz beam is collinearly overlapped with 8 fs NIR pulses at a variable time delay for electro-optic sampling as well as for precise beam alignment in the STM. A 300 µm thick ZnTe(110) crystal is used for electro-optic detection of the THz field outside UHV in a reference beam path that is identical to the STM beam path. The EOS signal is deconvoluted with the detector response in the time domain[22,43] to obtain the original THz electric field (EOS signals and deconvoluted fields are shown in the Supporting Information).

Numerical simulations are performed to calculate the THz near-field in the tip-sample junction by solving the time-harmonic wave equation for the electric field within the RF-Module of COMSOL Multiphysics 5.5. Details are provided in the Supporting Information.

…………

## AUTHOR INFORMATION

**Corresponding Author**

*E-mail: m.mueller@fhi-berlin.mpg.de

**Notes**

The authors declare no competing financial interest.

## Acknowledgement

The authors thank T. Kumagai, A. Paarmann and F. Krecinic for valuable discussions. The authors further thank T. Kumagai, Unisoku Inc. and M.B. Raschke for support and discussions in the development of the STM instrumentation. We thank G. Jakob and M. Kläui for providing us with the spintronic THz emitter, and H. Kirsch for tip preparation support. N. M.-S. and T.K. are grateful for funding through the ERC H2020 CoG project TERAMAG/Grant No. 681917. T.K. and M.W. acknowledge the German Research Foundation (DFG) for support through project B05 and B07 of the SFB/TRR227 Ultrafast Spin Dynamics.

**Supporting Information.** Supporting Information Available: Discussion of space charge effects, dependence of THz waveforms on DC bias and THz voltage, antenna model, measurement of incident THz electric fields, dependence of THz spectra and amplitude on STE position, estimation of tip-sample distance at the setpoint, numerical simulation of THz electromagnetic field distributions. This material is available free of charge via the Internet at http://pubs.acs.org.

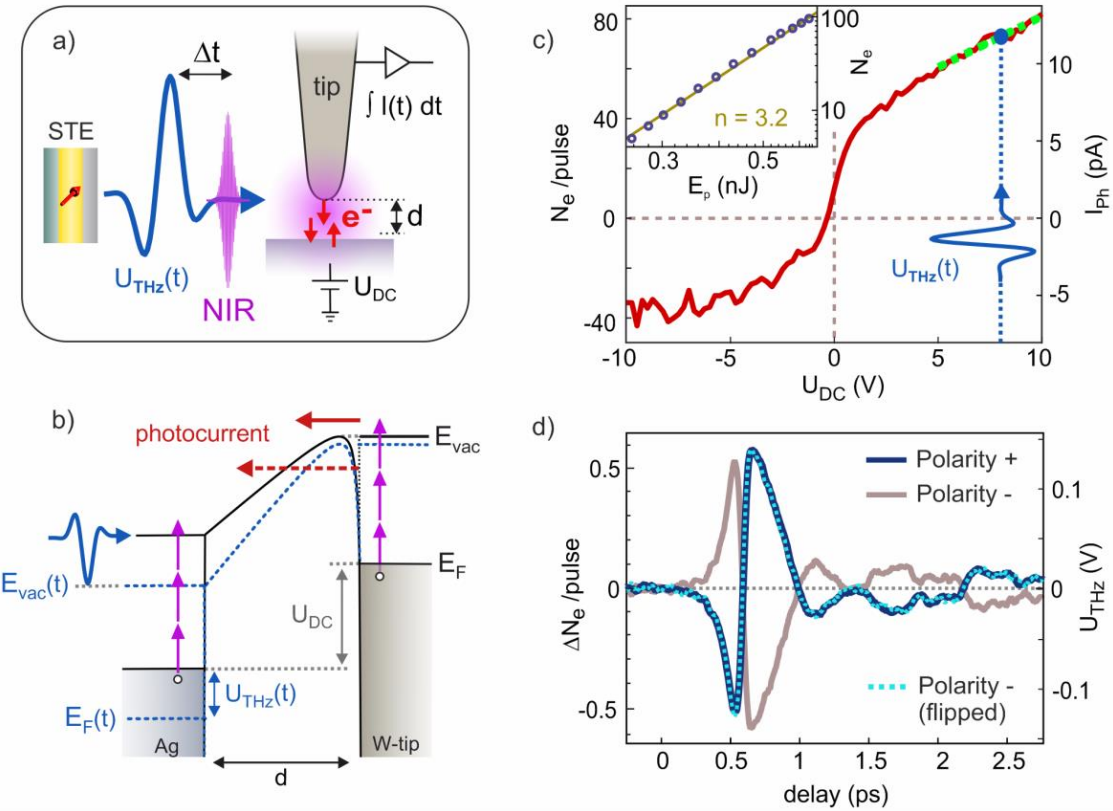

**Figure 1**: (a) Experimental scheme and (b) junction diagram for sampling of the THz voltage transient applied to the STM. The THz electric field emitted from the spintronic emitter (STE) acts as an instantaneous bias which modulates the photocurrent generated by 8 fs NIR laser pulses. Depending on the barrier height and distance $d$, the photocurrent originates from above-barrier photoemission and/or photo-assisted tunneling through the barrier. (c) Laser-induced photocurrent-voltage characteristics and (d) tip-enhanced THz waveforms sampled at 8 V DC bias ($d = 1 \ \mu m$, $E_{p,STE} = 0.6 \ \mu J$) for positive and negative THz polarity (STE magnetization M+/M-). The THz voltage (right y-axis) is retrieved from the THz-induced photocurrent change (left y-axis) via the linear I-V slope (green dashed line in c).The power law scaling of the photocurrent (inset in (c), $U_{DC} = 8$ V) reveals operation in the multiphoton photoemission regime.

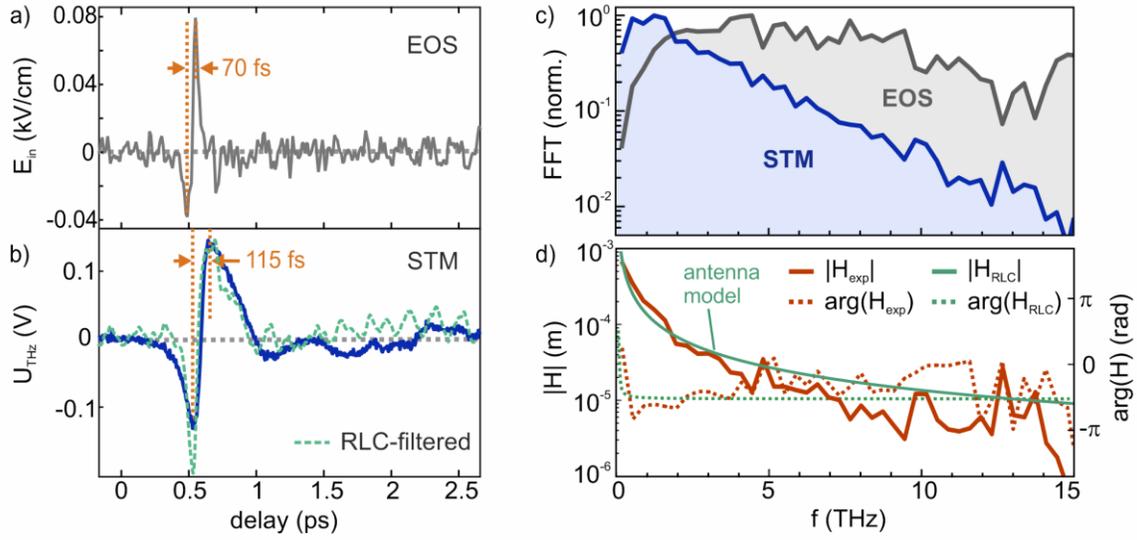

**Figure 2**: Comparison of (a) incident THz electric field and (b) tip-enhanced THz voltage transient in the STM junction ($U_{\mathrm{DC}} = 8$ V, $d = 1$ μm, $E_{\mathrm{p,STE}} = 0.6$ μJ). (c) THz amplitude spectra of the waveforms of (a) and (b), revealing strong low-pass filtering from the tip, which is characterized by (d) the measured receiving antenna transfer function H(ω). The green line in (d) shows the response of the antenna model with $R = 300$ Ω, $L = 0.32$ nH and $C = 35$ fF, yielding a resonance frequency of 0.05 THz and the *RLC*-filtered voltage transient plotted in b) (dashed curve).

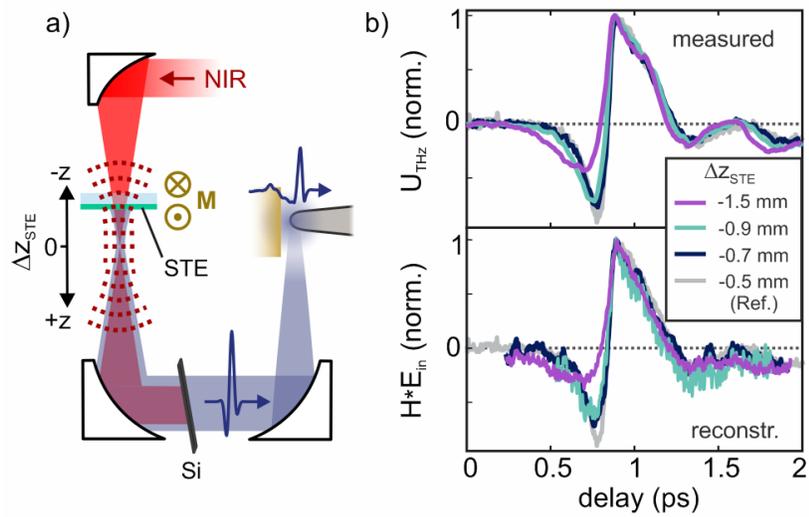

**Figure 3:** (a) Curved-wavefront excitation geometry with varying positions $\Delta z_{STE}$ of the STE through the convergent NIR pump beam. (b) THz waveforms for various values $\Delta z_{STE}$ close to the NIR focus. The waveforms measured in the STM (top panel) are reproduced by the waveforms obtained from inverse Fourier transformation (bottom panel) of the product of the measured tip transfer function $H(\omega)$ (Fig. 2d) and the respective incident THz spectra (shown in the Supporting Information). The light grey curve shows the reference waveform from Figure 2 used to obtain $H(\omega)$. All other settings are the same as in Figure 2.

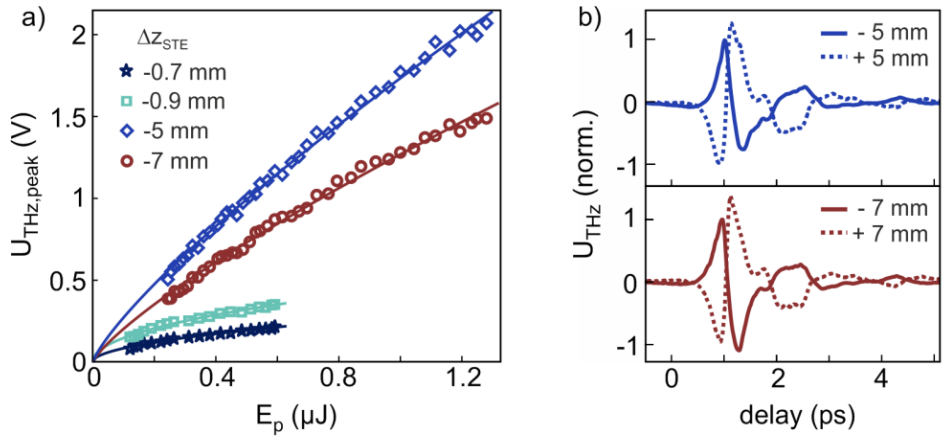

**Figure 4**: (a) Peak THz voltage vs STE pump pulse energy for different positions of the STE ($\Delta z_{STE}$) in front of the NIR focus. The THz voltage scales sublinearly with the pump pulse energy due to thermal saturation depending on the NIR spot size and hence STE position. Lines are sublinear fits to the data. At very small spot sizes close to the NIR focus ($|\Delta z_{STE}| < 1$ mm) ablation occurs and prevents excitation at high pulse energies in the current setup. (b) THz voltage transients in the STM for two positions $\Delta z_{STE} = \pm 5$ mm and $\Delta z_{STE} = \pm 7$ mm of the STE far away from the NIR spot size. The dashed lines show the corresponding waveforms for the STE positioned at the same distance in the divergent NIR beam. ($U_{DC} = 8$ V, $d = 1$ μm)

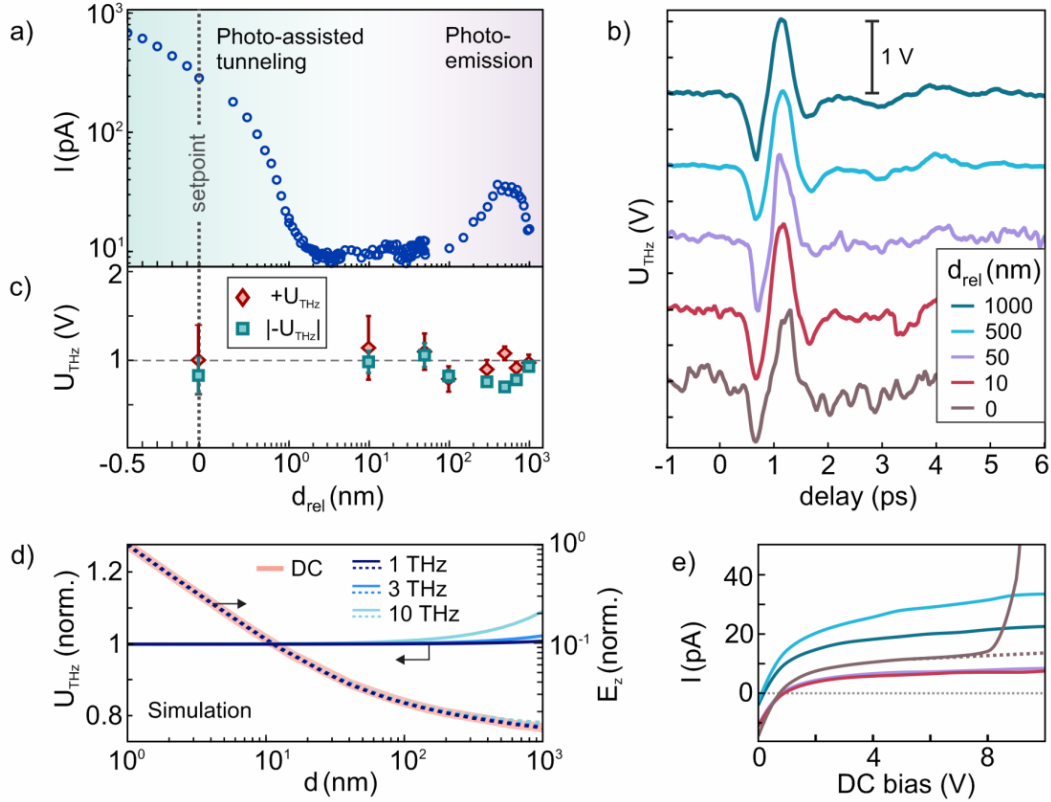

**Figure 5**: Dependence of the (a) photocurrent, (b) tip-enhanced THz waveform and (c) peak THz voltage of the positive and negative half cycle on the tip-sample distance relative to a setpoint of $I = 200$ pA and $U_{DC} = 10$ V. Nearly identical THz waveforms and constant THz voltages are observed. d) Numerical simulation of the THz potential difference induced between tip and sample (blue solid lines, left y-axis). The THz electric field (blue dashed lines) and DC electric field (light-red solid line) in the STM junction are plotted at the right y-axis ($E_z$ is taken 0.1 nm below the tip center, plotted values are normalized). e) I-V curves used for THz voltage calibration (waveform sampling performed at $U_{DC} = 8$ V). At $d_{rel} = 0$ nm the dashed line is used for calibration, which is the photocurrent I-V without background from DC field emission, which is assumed to exponentially increase with bias.

Supporting Information:

# Phase-resolved Detection of Ultrabroadband THz Pulses inside a Scanning Tunneling Microscope Junction

Melanie Müller[1]*, Natalia Martín-Sabanés[1,2], Tobias Kampfrath[1,2], and Martin Wolf[1]

[1]*Department of Physical Chemistry, Fritz-Haber Institute of the Max-Planck Society, Faradayweg 4-6, 14195 Berlin, Germany.*

[2]*Department of Physics, Freie Universität Berlin, Arnimallee 14, 14195 Berlin, Germany.*

*Corresponding author: m.mueller@fhi-berlin.mpg.de

*10 pages, 5 figures*

1. **Discussion of space charge effects**

2. **Dependence of THz waveforms on DC bias and THz voltage**

3. **Antenna model**

4. **Measurement of incident THz electric fields**

5. **Dependence of THz spectra and amplitude on STE position**

6. **Estimation of tip-sample distance at the setpoint**

7. **Numerical simulation of THz electromagnetic field distributions**

# 1. Discussion of space charge effects

Sampling the THz voltage across the STM junction via THz-induced modulation of the photocurrent requires that the photocurrent reacts quasi-instantaneous on the time scale of the applied THz field. However, several effects such as photoelectron propagation in the tip-sample gap or long-lived hot carriers in the photoexcited tip or sample may disturb the measurement and have to be ruled out. As discussed by Yoshida et al.,[1] THz-Streaking back into the tip can be circumvented by operating at high DC bias and low THz fields[1], whereas lifetime effects from hot carriers can be neglected when operating in the regime of multiphoton photoemission above the potential barrier. Another effect, not considered by Yoshida et al.,[1] is space charge, i.e., the Coulomb repulsion experienced by electrons in a photoelectron cloud containing multiple electrons. Due to the nanometer-sized volume at the tip apex, even a low number of electrons can lead to high photoelectron densities, leading to strong Coulomb interaction between the photoelectrons. As a consequence, 'leading' photoelectrons at the front of the space charge cloud are accelerated away from the tip faster than by the DC acceleration alone, whereas 'backside' photoelectrons close to the tip surface are accelerated back towards the tip by the space charge cloud in front of them. Those back-accelerated electrons can then be steered back into the tip by THz fields much lower than the DC field due to the described effect of space charge acceleration. Moreover, these photoelectrons are not rapidly accelerated away from the tip as expected from the large inhomogeneous DC field at the tip, but can remain a significant amount of time close to the tip surface. Hence, they not only experience the instantaneous THz field at the time of photoemission, but can be steered back into the tip by a THz field at a much later delay within the THz pulse, i.e., the overall process is not instantaneous and can act as an effective low pass filter distorting the measured THz waveform.

---

[1] It should be noted that, even at low THz and high DC fields, the THz field always affects the photoelectron trajectories in the tip-sample gap, depending on the emission time and flight time of the photoelectrons with respect to the arrival and duration of the THz pulse. As we detect the total current and average over all energies and arrival times of the photoelectrons, our measurement is not sensitive to small modulations of the electron trajectories, but is only sensitive to a reduction of the total electron yield by THz-streaking back into the tip.

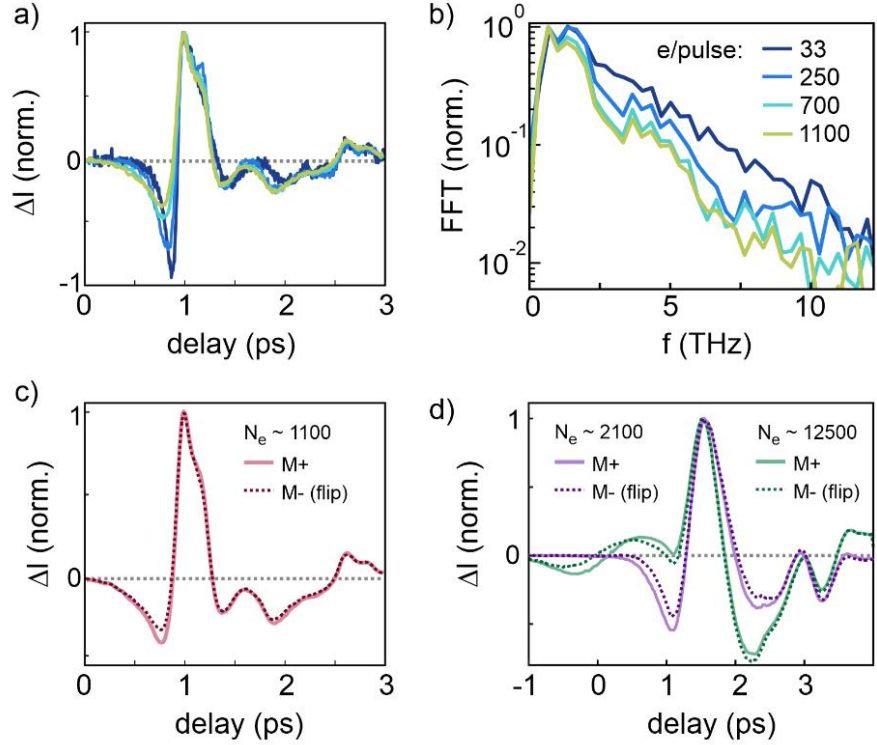

Figure S1. (a) Tip-enhanced THz waveforms and (b) corresponding Fourier amplitude spectra measured in the STM junction for increasing numbers of photoelectrons emitted per NIR pulse (corresponding increasing pulse energies $E_{p,STM}$ are 0.45 nJ, 0.95 nJ, 1.42 nJ and 1.8 nJ). Large electron numbers/densities lead to low-pass filtering presumably due to space charge. In the presence of space charge, the THz waveforms for opposite THz polarities exhibit different shapes, as shown in (c) for 1100 e/pulse ($E_{p,STM}$ = 1.8 nJ) and (d) for 2100 e/pulse ($E_{p,STM}$ = 4 nJ) and 12500 e/pulse ($E_{p,STM}$ = 12.5 nJ). At very high electron density, severe waveform distortions occur. We note that (d) has been measured for a different tungsten tip. ($d$ = 1 μm, $U_{DC}$ = 8 V, $E_{p,STE}$ = 0.4 μJ, $U_{THz,peak}$ = 0.12 V)

In Figure S1 we plot THz waveforms measured in the STM junction at 1 μm gap distance and 8 V DC bias for different photocurrents. The applied THz voltage is 0.12 V at the maximum peak. As can be seen from Figure S1a), the measured waveform clearly depends on the number of electrons excited per laser pulse, with a most pronounced deformation at the beginning of the pulse before the main half cycle. This is accompanied by a reduction of the THz bandwidth as plotted in Figure S1b), revealing the effective low pass filtering due to space charge. Moreover, Figure S1c) shows that in this regime the shape of the THz waveform also depends on the THz polarity. These deformations and deviations with polarity are most pronounced at the beginning of the THz pulse. This is not surprising as those photoelectrons excited shortly before the arrival of the main pulse experience the strongest

THz field, whereas the photoelectrons excited at later times only experience the ringing oscillations after the main THz cycle. For even higher electron numbers, severe waveform distortions are observed as plotted in Figure S1d), with distortions and polarity dependence occurring over the full THz waveform.

Understanding the exact waveform deformations in the presence of space charge requires a detailed analysis of the dynamics of the photoelectron cloud in the tip-sample gap and is beyond the scope of this work. At this point, we conclude that elimination of space charge contributions is crucial for reliable THz waveform sampling by the employed technique as described in the main manuscript. In our experiments we proof the absence of space charge effects to our measurements by systematically reducing the number of photoelectrons to a value at which no waveform distortions and no spectral filtering is observed by further lowering the photocurrent. These observations point out the importance for a careful consideration of non-instantaneous effects potentially disturbing the sampled THz voltage transient.

## 2. Dependence of THz waveforms on DC bias and THz voltage

The propagation of photoelectrons in the tip-sample gap (and potentially back to the tip) depends on the combined forces of the DC field, the THz field, and space charge interactions. Moreover, waveform distortions from photo-assisted tunneling of electrons close to the Fermi level with longer lifetimes may be expected to contribute at high DC bias and short gap distances. As the lifetime of hot electrons depends on the energy window sampled by the THz pulse, distorting effects from hot carriers should also depend on the DC bias and THz voltage amplitude. Hence, undistorted instantaneous THz near-field sampling requires that the measured THz waveform does not depend on the DC bias and incident THz field strength within the linear range of the $I_{Ph}$-V slope (see Figure 1c)). Figure S2b) shows THz waveforms measured for three different DC biases with the corresponding $I_{Ph}$-V curve plotted in Figure S2a). All waveforms exhibit the same shape and voltage amplitude as expected from the linear slope of the I-V curve. Figure S2c) shows THz waveforms measured for two incident THz amplitudes at 10 V bias, and again we find that the sampled THz waveform does not depend on the applied THz voltage. In addition to the constant THz waveform observed for opposite THz polarities, as plotted in Figure 1d) in the main manuscript, these results confirm the interpretation and validity of our sampling approach.

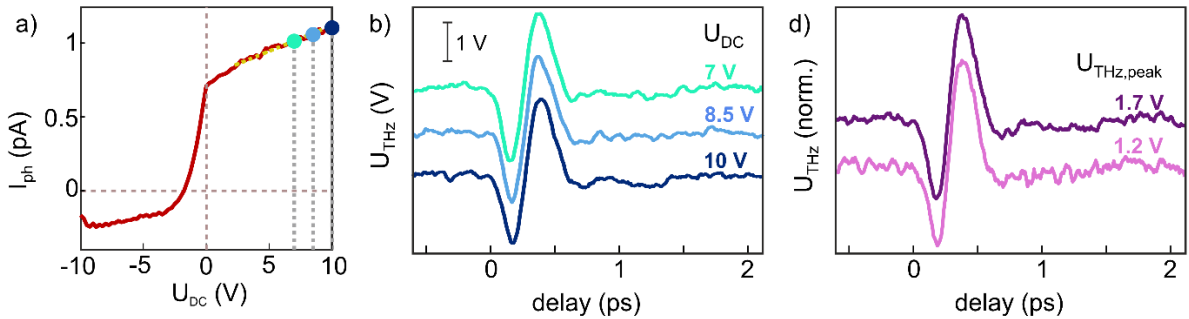

*Figure S2. (a) I-V curve used for calibration of the tip-enhanced THz waveforms shown in (b) and (c). THz waveforms are measured in the STM at (b) different DC biases ($E_{p,STE} = 0.9\ \mu$, $U_{THz,peak} = 1.9\ V$) and (c) different THz amplitudes ($U_{DC} = 10\ V$, $E_{p,STE} = 0.48\ \mu$ and $E_{p,STE} = 0.9\ \mu J$) in the linear part of the I-V slope. ($d = 1\ \mu m$, STE position $\Delta z_{STE} = -2.6\ mm$). These data were measured with a Ag-tip sharpened by a focused ion beam milling. Similar results are obtained from tungsten and Pt/Ir tips.*

## 3. Antenna model

The model used to fit the tip antenna response in Figure 2 treats the tip as an *RLC* electronic circuit with resistance *R*, capacitance *C* and inductance *L* connected in series. In this very simple model, the incident THz electric field applies a voltage to the antenna that leads to a current induced inside the antenna of

$$I_{\text{THz}}(\omega) \propto \frac{E_{\text{in}}(\omega)}{R + i\omega L - i(\omega C)^{-1}} \ .$$

Best reproduction of the THz voltage transient from the incident THz waveform is obtained for circuit parameters of $R = 300\ \Omega$, $L = 0.32$ nH and $C = 35$ fF, indicating an effective antenna length of ~1 mm for a 300 μm thick wire. These values are in good agreement with previously reported results[2,3]. The simple *RLC* model considers only the tip wire alone as the antenna. A more detailed theoretical analysis of the receiving properties of an STM junction requires more advanced models taking into account the dielectric response of the sample, the exact STM environment, and scattering e.g. at the tip shaft or at the tip mount.

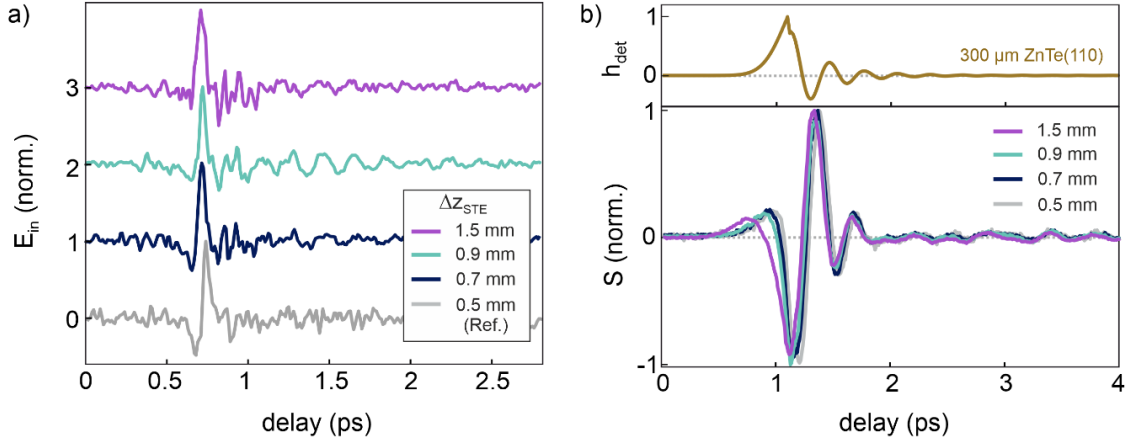

*Figure S3. (a) THz electric field transients incident onto the ZnTe(110) detection crystal for four different STE positions as obtained from deconvolution of the EOS signal with the detector response. The phase shift applied by moving the STE inside the focused NIR beam is apparent in the waveforms. (b) Corresponding EOS signals (bottom) and detector response function $h_{det}(t)$ (top).*

## 4. Measurement of incident THz electric fields

To analyze the THz electric field incident to the STM junction, we pick the THz beam before entering the STM chamber and focus the THz pulses in a 300 µm thick ZnTe(110) crystal for electro-optic sampling. The path length and optical components are identical to the STM beam path. In EOS, the measured time-domain signal $S(t)$ is given by the convolution of the THz electric field $E_{in}(t)$ incident on the detector with the detector response function $h_{det}$,

$$S(t) = (h_{det} * E_{in})(t).$$

Hence, if $h_{det}(t)$ is known, the THz electric field can be obtained by deconvolution of the EOS signal with the detector response, which depends on the properties and thickness of the electro-optic medium and the sampling pulses. After calculation of $h_{det}(t)$, the deconvolution is performed numerically as described in more detail in references[4,5]. Figure S3a) shows the deconvoluted waveforms of the THz electric field incident on the STM tip for the four STE positions plotted in Figure 3b). As mentioned in the manuscript, the measured EOS amplitude at frequencies >5 THz is comparably low, leading to increased noise at higher frequencies in the broadband deconvolution process. This can be mitigated by using thinner detection crystals. However, echos in the first picoseconds caused by reflections inside the detection crystal can complicate the analysis of the transfer function in this case.

Direct comparison of the THz electric field obtained from EOS and the THz voltage waveform measured in the STM requires that both waveforms are recorded at the same position along the focused THz beam. In the EOS setup, the detection crystal is positioned in the focus of the tightly focused NIR beam that propagates collinear with the THz beam and is used as sampling pulse. Likewise, we use the same collinearly propagating NIR pulses as alignment beam to adjust the THz focus on the tip in the STM. Precise alignment is hereby ensured by monitoring the light reflected off the Ag sample and collected via the second (Ag) parabolic mirror inside the STM chamber, and by optimization of NIR-induced photoemission from the tip apex.

## 5. Dependence of THz spectra and amplitude on STE position

As discussed in the main manuscript, the THz spectra can vary slightly with STE position. Figures S4a) and S4b) show the THz amplitude spectra of the tip-enhanced THz waveforms plotted in Figures 3b) and 4b), respectively. We observe a slight bandwidth reduction upon moving the STE away from the NIR focus. We assign such variation of the THz spectrum to the imperfect imaging of the THz focus onto the tip at STE positions outside the NIR focal plane. The variation of the peak THz voltage amplitude with STE position $\Delta z_{STE}$ is shown in Figure S4c). Maximizing simultaneously the THz amplitude and bandwidth requires a modified STE geometry optimized for our laser parameters, which is currently under investigation.

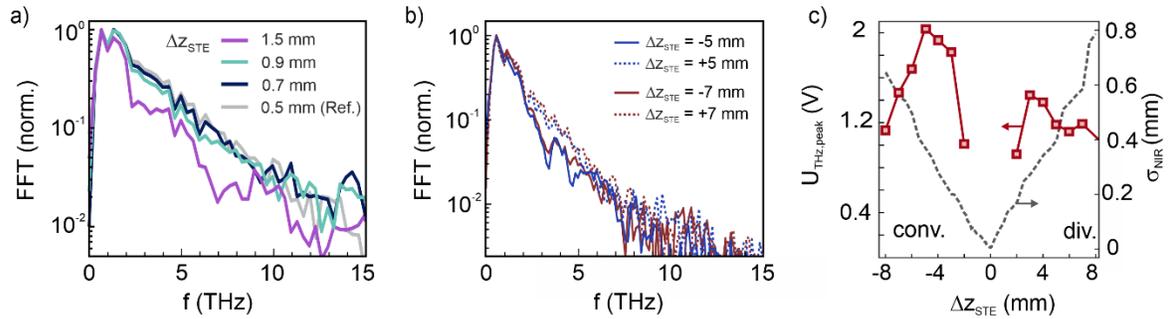

Figure S4. Fourier amplitude spectra of the tip-enhanced THz waveforms vs STE position shown (a) in Figure 3b), and (b) in Figure 4b). We observe a red-shift of the spectra and slightly reduced THz bandwidth when moving the STE away from the focus. (c) Peak THz voltage as a function of STE position ($E_{p,STE}$ = 1.2-μJ). The black dashed line shows the NIR pump spot size vs STE position (right y-axis).

## 6.  Estimation of tip-sample distance at the setpoint

The actual tip-sample distance under NIR illumination is estimated as follows: First, we estimate the change in gap distance due to removal of the photoinduced current contribution by measuring the prompt (~30 ms, which is the time required to reach again the setpoint current) reduction in gap distance upon blocking the NIR beam, which gives $\Delta d_{\text{ph}} \cong 1.6$ nm. Second, we estimate the absolute distance of the 'dark' junction by measuring the z-dependence after thermal equilibration and extrapolation of the I-z-curve to the quantum conductance $G_0$. This yields a tip-sample distance of $\Delta d_{\text{DC}} \cong 3$ nm at 8 V without illumination. We obtain the same distance $\Delta d_{\text{DC}}$ by extrapolating I-z-curves at 1 V set bias (yielding 1 nm gap distance) and observation of the change of the z-piezo position by manually changing the set bias from 1 V to 8 V during feedback (which gives ~2 nm). The absolute distance of the illuminated junction is then estimated to be the sum of the two contributions, $d = \Delta d_{\text{DC}} + \Delta d_{\text{ph}} = 4.6$ nm. Although this procedure neglects possible influences of the tip temperature on the DC current, it gives a reasonable estimate of the actual tip-sample distance.

## 7.  Numerical simulation of THz electromagnetic field distributions

Numerical simulations are performed to calculate the THz response of the STM tip by solving the time-harmonic wave equation for the electric field within the RF-Module of COMSOL Multiphysics 5.5. For a given excitation frequency, COMSOL solves for the full time-harmonic electromagnetic field distribution. Static field distributions are calculated using the AC/DC module of COMSOL. Simulations are performed in 3D with the tip oriented along the z-axis. The tip is modelled by a conical wire with a half opening angle of 4° terminating in an apex of 50 nm radius. The sample of 10 µm thickness is placed at a variable distance $d$ in front of the tip. The width of the simulation volume is 200×200 µm and the wire is truncated by the end of the simulation volume at a height of 200 µm from the apex. The simulation volume is surrounded by perfectly matched layers to absorb all outgoing waves.

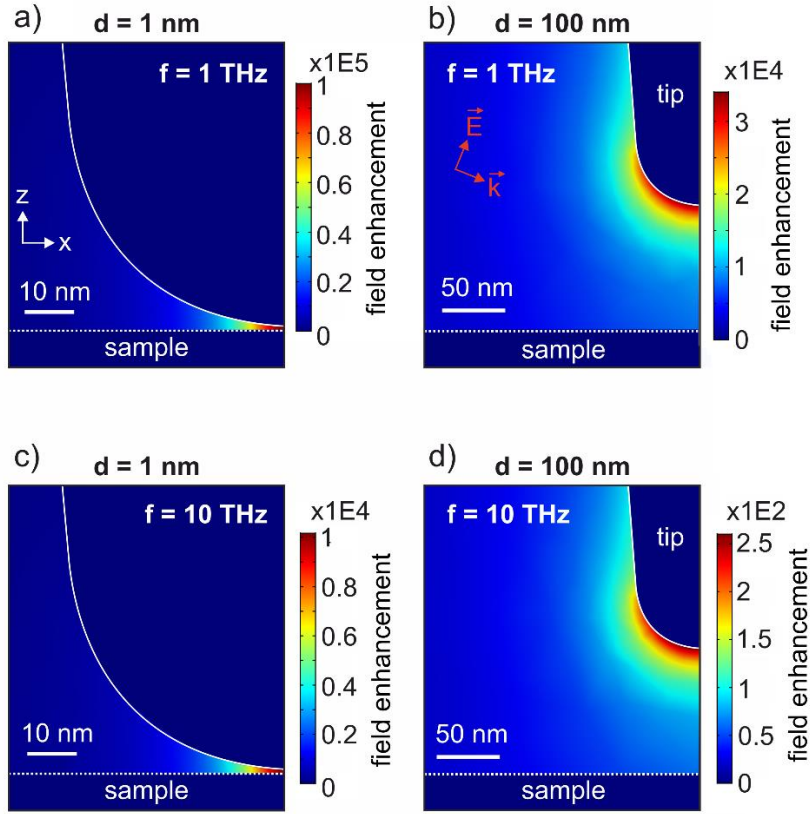

*Figure S5.* *Spatial distributions of the tip-enhanced THz electric field (normalized field) in the tip-sample junction at 1 nm and 100 nm gap distances for frequencies of (a-b) 1 THz and (c-d) 10 THz, respectively. The color bars represent the enhancement of the THz field inside the junction with respect to the incident THz field amplitude. The red arrows indicate the direction and polarization of the incident THz field. Simulation parameters are described in the text.*

The tip-sample junction is illuminated by plane-wave THz radiation propagation along the x-direction at an angle of 68° with respect to the tip axis as and is polarized linearly in the $x$-$z$-plane, as indicated by the red arrows in Figure S5b). The simulation volume is cut in half along the y-direction (out of plane) according to the symmetry given by the THz beam direction to reduce the computational cost. The materials properties of the tip and sample are determined from the complex dielectric functions of tungsten and silver, whose real and imaginary parts are calculated from the Lorentz Drude model with the Lorentz-Drude parameters taken from Rakic et al.[6].

Figures S5a)-d) show spatial distributions of the tip-enhanced THz electric field in the tip-sample junction for two THz frequencies. As can be seen from the color scales, the THz field

enhancement increases significantly with lower THz frequencies and shorter gap distances. At far distances b) and d), the strongly inhomogeneous THz near-field is confined to the tip apex with a spatial extent given by the tip radius, and decays rapidly with larger distances away from the tip surface. At nanometer distances much smaller than the tip radius, the field becomes spatially more localized inside the tip-sample gap, and becomes homogeneous like in a plate-capacitor at the junction center $(x, y) = (0,0)$. We obtain the THz-field-induced potential difference from the computed fields by line integration of the z-component of the THz near-field along the center of the junction at $(x, y) = (0,0)$. We confirmed that line integration along other pathways yields the same THz-induced potential difference.